\documentclass[aps, prd, reprint, showpacs, floatfix, letterpaper,
 nofootinbib, superscriptaddress,longbibliography]{revtex4-2}
\usepackage[version=4]{mhchem}
\usepackage{blindtext}
\usepackage[newitem]{paralist}
\usepackage{graphicx}
\usepackage{epsfig}
\usepackage{bm}
\usepackage{amsfonts}
\usepackage{pgf}
\usepackage{color}
\usepackage[T1]{fontenc}
\usepackage[utf8]{inputenc}
\usepackage{graphicx}
\usepackage[english]{babel}
\usepackage{amsmath}
\usepackage{amssymb}
\usepackage{amsfonts}
\usepackage{makecell}
\usepackage[draft,deletedmarkup=xout]{changes}
\usepackage{mathtools}
\usepackage{xcolor}
\usepackage{multirow}
\usepackage{mhchem}
\usepackage{wrapfig}
\usepackage{lipsum}
\usepackage{xcolor}
\usepackage{multirow}
\usepackage{mhchem}
\usepackage{hyperref}
\usepackage{dcolumn}

\definecolor{green}{rgb}{0.19,0.64,0.54}
\definecolor{blue}{rgb}{0,0,1}
\definecolor{reddish}{rgb}{0.65, 0.2, 0.2}
\definecolor{darkgreen}{rgb}{0.2,0.7,0.3}
\definecolor{darkblue}{rgb}{0.3,0.40,0.48}
\definecolor{gray}{rgb}{.8,.8,.8}

\hypersetup{
	pdftitle={Cosmic String Networks with Time-Varying Tension},
  	pdfauthor={Jos\'{e} Correia, and Ivan Rybak},
    colorlinks=true,       
    linkcolor=darkblue,          
    citecolor=darkgreen,        
    filecolor=reddish,      
    urlcolor=blue,          
    linktocpage=true
}

\begin{document}

\title{Cosmic String Networks with Time-Varying Tension: \texorpdfstring{\\}{ }Field-Theory Simulations and an Analytic Model}

\author{Jos\'{e} Correia}
\email{jose.correia@helsinki.fi}
\affiliation{Institute of Theoretical Astrophysics, 
University of Oslo, P. O. Box 1029 Blindern, N-0315 Oslo,Norway}

\author{Ivan Rybak}
\email[]{irybak@ibs.re.kr}
\affiliation{Cosmology, Gravity, and Astroparticle Physics Group, Center for Theoretical Physics of the Universe,
Institute for Basic Science (IBS), Daejeon 34126, Korea}

\begin{abstract}
Cosmic string networks are predicted in some early-Universe scenarios, and in certain models, the string tension becomes a function of time. Here, for the first time, we perform numerical field-theory simulations of cosmic string networks with time-varying tension in an expanding Universe. We use a modified Abelian-Higgs model in which the tension varies as a power law of the scale factor $a$, that is, $\mu \propto a^{-2m}$. We perform simulations on grids containing $8192^3$ grid points, with $m\in[-0.5,0.5]$, during radiation- and matter-dominated epochs. We find that, in our model, a string network forms only when the potential decreases more slowly than the amplitude of field oscillations redshifts due to the expansion of the Universe. Once formed, the network evolves according to the standard scaling behavior predicted by the velocity-dependent one-scale (VOS) model. To interpret these results, we compare the predictions of the VOS model with time-dependent tension to the asymptotic values obtained from the numerical simulations. Overall, we find good agreement between the numerical field-theory simulations and the calibrated VOS model. These results shed light on the evolution of cosmic string networks with time-varying tension and have potential implications for their observational signatures.
\end{abstract}

\date{\today}
\maketitle

\section{Introduction}
\label{I}

Cosmic strings are intriguing hypothetical objects, the existence of which was originally proposed in~\cite{Kibble:1976sj}. Many extensions of the Standard Model of particle physics predict the creation of such objects in the early stages of the Universe \cite{Hindmarsh:1994re, Vilenkin:2000jqa, Jeannerot:2003qv, Copeland:2009ga,Kibble:1982ae,Dror:2019syi,Vilenkin:1982ks}. If cosmic strings were ever created in the early stages of the Universe, they would leave distinctive observational features in cosmological backgrounds \cite{SantanaMota:2014xpw, Lizarraga:2016onn, Bulygin:2023cqq, Baeza-Ballesteros:2024sny,Hindmarsh:2022awe,Auclair:2019wcv,Planck:2013mgr,NANOGrav:2023hvm}.

To make accurate predictions of observational signals from cosmic strings, we need to understand the dynamics of string networks in the expanding Universe. This task has been approached from different directions. One of them is to perform field-theory simulations. The most well-tested approach to cosmic string field-theory simulations is based on the Abelian-Higgs (AH) model with critical coupling, as described in Refs.~\cite{Moore:2001px, Bevis:2006mj, Correia:2021tok}. This is the most consistent approach; however, it lacks the ability to satisfy the real cosmological hierarchy of scales, namely, to resolve a string core that is multiple orders of magnitude smaller than a box the size of the observable Universe (unless circumvented by unphysical modifications of the equations of motion \cite{Press:1989yh,Bevis:2006mj}). On the other hand, another approach, less susceptible to resolution issues, was also studied, under the so-called Nambu-Goto (NG) approximation \cite{Goto:1971ce,Nambu:1970,Forster:1974ga}. Here cosmic strings are treated as objects with zero thickness \cite{Martins:2005es, Ringeval:2005kr, Blanco-Pillado:2011egf,Allen:1990tv,Bennett:1989yp}. Despite their differences (an open problem in the field \cite{Blanco-Pillado:2023sap,Baeza-Ballesteros:2024sny,Matsunami:2019fss,Aurrekoetxea:2026xrl}), both types of simulations approach a self-similar scaling regime in the expanding Universe, i.e., the string energy density decreases as $t^{-2}$ and the root-mean-square (rms) velocity remains constant.

Although the evolution of a string network is a highly nonlinear problem, the existence of scaling behavior was predicted analytically in Ref.~\cite{Kibble:1984hp}. A more accurate description of the evolution of a cosmic string network, known as the velocity-dependent one-scale (VOS) model, was later developed in Refs.~\cite{Martins:1996jp, Martins:2000cs}. Recently, it was demonstrated that, with the inclusion of phenomenological energy-loss parameters, the VOS model can reproduce the asymptotic scaling values obtained from field-theory simulations of the AH model at critical coupling \cite{Correia:2019bdl,Correia:2021tok}.

In this work, we consider a cosmic string model with time-varying tension, which was initially suggested in Ref.~\cite{Yamaguchi:2005gp}. The scenario of time-varying cosmic string tension has been considered in different contexts of high-energy physics: within the field-theory framework \cite{Emond:2021vts} and in the context of specific properties of cosmic superstrings \cite{Revello:2024gwa, Conlon:2024uob, Brunelli:2025ems, SanchezGonzalez:2025uco,  Brunelli:2025eif}. We additionally remark that significant attention has also been devoted to the existence of scaling for networks of axion strings (which also possess a time-dependent tension). This is a debated topic that significantly impacts axion mass predictions (see for instance \cite{Yamaguchi:1998gx, Kawasaki:2018bzv, Gorghetto:2018myk, Martins:2018dqg, Klaer:2019fxc,Hindmarsh:2021vih,Benabou:2024msj, Correia:2024cpk,Saikawa:2024bta}).

In this study, we introduce a varying string tension based on the standard AH field-theory model (Section~\ref{II}) and perform numerical simulations to verify the existence of a scaling regime (Section~\ref{IV}). We also analyze the obtained results within the VOS framework.

\section{Field-theory model of cosmic strings with varying tension}
\label{II}

We use the AH model, which is described by the following Lagrangian:
\begin{equation}
\label{LagrangIn}
    \mathcal{L} = \left|D_{\mu}\phi \right|^2 - \frac{1}{4 e^2} F^{\mu \nu} F_{\mu \nu} - \frac{\lambda}{4}\left(\left| \phi \right|^2 - \eta^2 \right)^2 ,
\end{equation}
where $\phi$ is a complex scalar field, $F_{\mu \nu} \equiv \partial_{\mu} A_{\nu} - \partial_{\nu} A_{\mu}$ is the field strength tensor, $D_{\mu} \equiv \partial_{\mu}-i A_{\mu}$ is the gauge-covariant derivative, $\mu, \, \nu = 0,1,2,3$ are indices of the space-time coordinates associated with the metric $g_{\mu \nu}$, and $\lambda$, $e$, and $\eta$ are positive constants.

The AH model, given by Eq.~\eqref{LagrangIn}, describes the spontaneous symmetry-breaking mechanism through which the vector field $A_{\mu}$ becomes massive after symmetry breaking \cite{Higgs:1964pj}. The masses of the vector and scalar fields in the vacuum are given by $m_{s} = \sqrt{\lambda} \eta$ and $m_v = \sqrt{2} e \eta$. If symmetry breaking in the AH model occurs in the early Universe, cosmic string production is expected to take place; see, e.g., the reviews in \cite{Vilenkin:2000jqa}. Here, we consider the model of strings with critical coupling, in which strings do not experience long-range forces. This occurs when the Bogomol'nyi ratio of the masses is equal to one, i.e., $\beta \equiv \frac{m_{s}}{m_v} = \frac{\sqrt{\lambda}}{e \sqrt{2}} =1$.

We consider the evolution of the string network in the background of a Friedmann–Lema\^{i}tre–Robertson–Walker (FLRW) metric, which is given by
\begin{equation}
    ds^2 = a^2 (d \tau^2 - d \textbf{x}^2),
\end{equation}
where $\tau$ is the conformal time and $a$ is the scale factor.

We introduce a variation in the string tension through the parameter $\eta$, i.e., we demand that $\eta \propto a^{-m}$, which, as we will show later, leads to a time-dependent string tension. We also want to keep the string core resolvable throughout the entire simulation keeping $\beta = 1$. These conditions can be satisfied if we postulate the following variations of the string parameters:
\begin{equation}
\begin{gathered}
\label{DynamParam}
    \eta = \eta_0 a^{-m}, \quad e = e_0 a^{-(1-s)}, \quad \lambda = \lambda_0 a^{-2(1-s)}.
\end{gathered}
\end{equation}
The dynamical parameters \eqref{DynamParam} of the potential lead to dynamical masses of the scalar and vector fields, namely,
\begin{equation}
\label{MassVariation}
   m_s , m_v \propto a^{-(1-s)-m}. 
\end{equation}

The equations of motion for the Lagrangian \eqref{LagrangIn} with dynamical parameters \eqref{DynamParam} can be written as (with the gauge condition $A_0 =0$)
\begin{equation}
\begin{gathered}
\label{EqOfMot}
     \ddot{\phi} + 2 \frac{ \dot{a} }{a} \dot{\phi} - D_i D_i \phi = - a^{2s} \frac{\lambda_0}{2} \left( \left| \phi \right|^2 - \eta_0^2 a^{-2m} \right) \phi, \\
     \dot{F}_{0 i} + 2(1-s) \frac{ \dot{a} }{a} F_{0i} - \partial_j F_{ji} = 2 a^{2 s} e_0^2 \text{Im} \left[ \phi^* D_{i} \phi \right].
\end{gathered}
\end{equation}
Gauss's law in that case is given by
\begin{equation}
    \partial_i F_{0 i} = 2 e_0^2 a^{2 s} \text{Im} \left[ \phi^* \dot{\phi} \right].
\end{equation}

One can show that the condition $m=s$ leads to a constant conformal cosmic string core size. In particular, substituting $\phi = \eta_0 a^{-m} f(\rho) \text{e}^{i n \theta}$ and $A_{\theta} = n (\alpha(\rho)+1)$ into Eq.~\eqref{EqOfMot}, where $\rho \equiv r m_{s0}$ and $t \equiv \tau m_{s0}$, with the index ``$0$'' indicating that the corresponding mass is evaluated at the initial time, one can obtain
\begin{equation}
\begin{gathered}
\label{EqOfMotProfile}
       \frac{ \partial_{\rho} \left( \rho \partial_{\rho}f   \right)}{ \rho } - \frac{n^2 f \alpha^2}{\rho^2} =  \frac{a^{2(s-m)} \left(  f^2 - 1 \right) f}{2} , \\
      \partial^2_{\rho} \alpha - \frac{ \partial_{\rho} \alpha }{r} = -\frac{ a^{2 (s-m)} \alpha f^2}{\beta_0^2} ,
\end{gathered}
\end{equation}
where, for $m=s$, we recover the standard cosmic string profile. In this case, the comoving string-core thickness, which is inversely proportional to the scalar and vector masses in conformal coordinates, remains constant, namely,
\begin{align}
\label{StingCore}
    & r_{s} \propto \frac{1}{a \, m_s} \propto \frac{1}{a \, m_v} \propto \frac{1}{a^{s-m}}.
\end{align}

The string tension becomes time-dependent and evolves as \cite{Vilenkin:2000jqa}
\begin{equation}
\label{TensionVariation}
\mu \approx 2 \pi \eta^2 h\left(\beta\right) \propto a^{-2m},   
\end{equation}
where $h\left(\beta\right)$ is a slowly varying function with $h(1) \approx 1$.

\subsection{Simulation set up}
\label{IIA}

We will use the discretization procedure of \cite{Correia:2019bdl,Correia:2021tok,Hindmarsh:2017qff} to evolve all field content of the simulation in a comoving grid of size $L=N \delta x$, where $N=8192$ and $\delta x = 0.5$, updating them in steps of conformal time $\delta \tau = 0.1$ (satisfying the Courant–Friedrichs–Lewy condition). Before evolving the fields on a comoving lattice, we must first set up the initial conditions. We begin at $\tau_{\rm ini}=50$ by setting $A_\mu = E_\mu = 0$ and $\Pi=0$, where $\Pi$ is the conjugate momentum of the scalar field. The scalar field $\phi$ is initialized as a Gaussian random field with $\langle \phi \rangle = 1$ and an amplitude determined by the power spectrum
\begin{equation}
P(|\textbf{k}|) = (\sqrt{2 \pi} l_\phi)^3 \text{e}^{-\frac{1}{2} (|\textbf{k}| l_\phi^2 )},
\end{equation}
where $l_\phi$ is the initial correlation length and $\textbf{k}$ is a wave-vector. Although varying this length allows us to study the approach to scaling, here we simply set it to $l_\phi=10$, following standard best practices for reaching the scaling regime with the smallest possible dynamical range \cite{Hindmarsh:2017qff,Correia:2024cpk}.

\begin{table}
    \begin{ruledtabular}
    \begin{tabular}{cc}
    Parameter & Value  \\
    \hline
    \rule{0pt}{3ex}
    $L$ & 4096 \\
    $\lambda_0$ & 2 \\
    $e_0$ & 1 \\
    $\eta_0$ & 1 \\
    $\tau_\text{ini}$ & 50.0 \\
    $\tau_\text{dif}$ & 70.0 \\
    $\tau_\text{final}$ & 2048.0 \\
    $a(\tau_\text{dif})$ & 1.0 \\
    $l_\phi$ & 10.0 \\
    $s=m$ & $-0.5,\ -0.25,\ 0.25,\ 0.5$
    \end{tabular}
    \end{ruledtabular}
    \caption{\raggedright A summary of the simulation parameters, ranging from the potential parameters $\lambda_0$, $e_0$ and $\eta_0$, to the initial diffusion $t_{\rm dif}$ and final times, the scale factor at the beginning of cosmological evolution $a(\tau_{\rm dif})$, the initial correlation length $l_\phi$, and the tension power-law index $m$.
    \label{tab:parameters}}
\end{table}

After this, large unphysical gradients are still present in the initial conditions, so we use gradient flow/diffusive evolution until $\tau_\text{dif}=70.0$ as in Refs.~\cite{Hindmarsh:2017qff, Correia:2020gkj}, with the physical values of $\lambda_0=2$, $e_0=1$, and $\eta_0=1$, resulting in a network of strings with a radius of $r_{s,0}=1/\sqrt{2}$ at the end of the diffusion process. Afterwards, we begin evolving the fields according to discrete versions of \eqref{EqOfMot} in an expanding background until half a light-crossing time $\tau_{\rm final}=2048$. 

Since setting $s = m$ implies a constant comoving radius, we can normalize the scale factor to unity at the end of diffusive period of evolution, thus ensuring that the string radius remains unchanged, $r_{s} = r_{s,0}$. We will run simulations in matter- and radiation-dominated epochs, i.e. where the scale factor behaves as $a \propto \tau^2$ and $a \propto \tau$, respectively. For each epoch, we will run $5$ simulations per value of tension power law $m$ each with a different random number seed, and we will use four $m$ values, contained in the range $m \in [-0.5,0.5]$. The total number of simulation runs is $40$, and each simulation took approximately $65$~minutes and was run using $1024$~Nvidia A100 graphics processing units on the Leonardo supercomputer \cite{Cineca2024Leonardo}. 

\begin{figure}[h]
\centering
\includegraphics[width=0.5\textwidth]{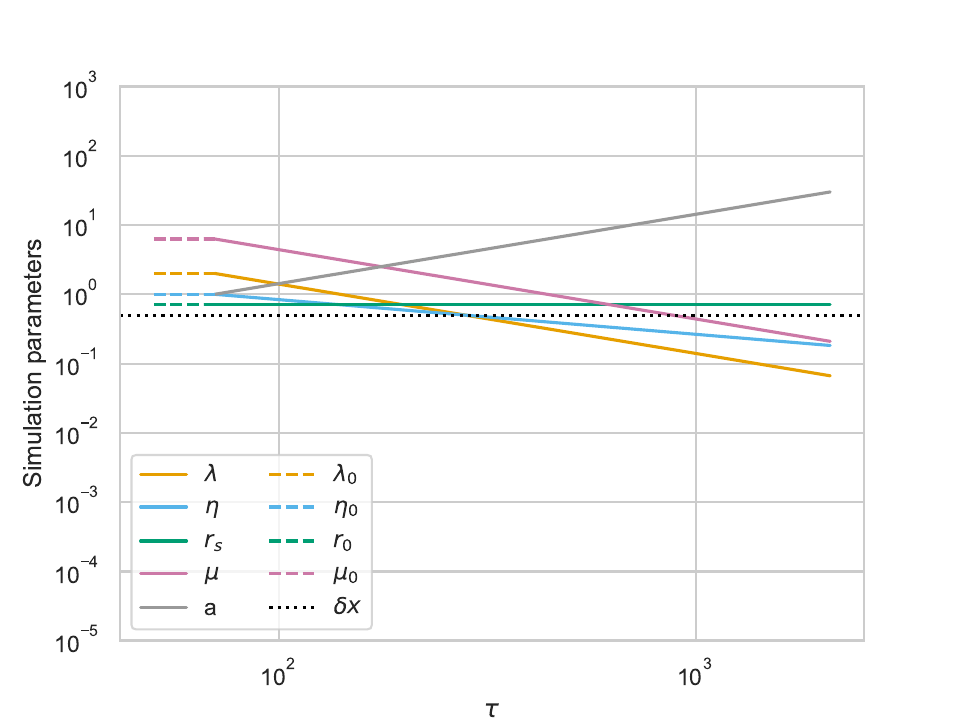}
\includegraphics[width=0.5\textwidth]{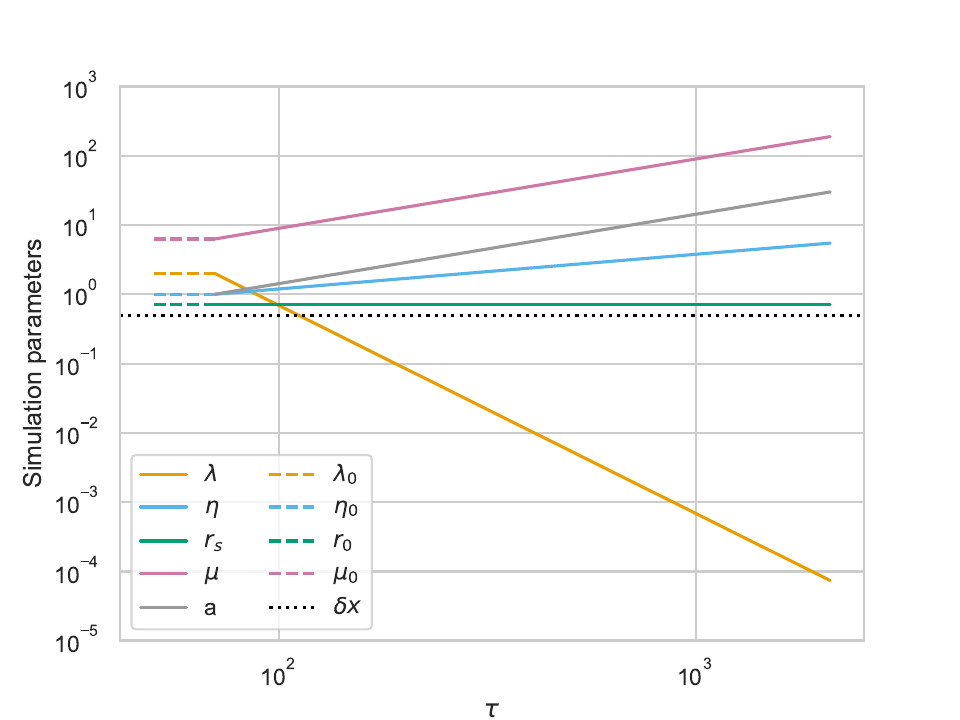}
\caption{Evolution of simulation parameters for diffusive evolution ($\lambda_0$, $e_0$, $r_{s,0}$) and in the radiation-dominated epoch ($\lambda$, $e$, $r$) along with lattice spacing $\delta x = 0.5$, scale factor $a(\tau)$ and symmetry breaking scale $\eta$ for an $8192^3$ simulation. The tensions during the diffusion, $\mu_0$, and during radiation, $\mu$, are also shown. The top panel shows the parameters for decreasing tension ($m=s=0.5$), while the lower panels show the parameters for increasing tension ($m=s=-0.5$).} \label{fig:params}
\end{figure}

A summary of the evolution of $\lambda$, $e$, $\eta$, string radius $r$ and scale factor $a$ can be found in Fig.~\ref{fig:params} for $m=s= \pm 0.5$ and a summary of all parameters is provided in Table~\ref{tab:parameters}.

\subsection{Observables: Length and Velocity}
\label{IIB}

We need to define a set of observables in simulations to fully characterize the evolution of the network. We begin by defining the energy density, Lagrangian density, and pressure in terms of their individual components
\begin{align}
\rho &= \rho_\pi + \rho_D + \rho_V + \rho_E + \rho_B, \\
\mathcal{L} &=  \rho_\pi -\rho_D - \rho_V + \rho_E - \rho_B, \\
p &= \rho_\pi -\frac{1}{3}\rho_D - \rho_V + \frac{1}{3}(\rho_E + \rho_B),
\end{align}
where each component is then defined as
\begin{align}
\rho_\pi &= |\pi|^2,  && \rho_D = |D_{i} \phi|^2,  \qquad \rho_E   = \frac{E^2}{2e^2 a^2}, \\
   \rho_B &= \frac{B^2}{2e^2 a^2}, && \rho_V   = \frac{\lambda}{4}(|\phi|^2 - \eta^2)^2,
\end{align}
where $E^i \equiv F^{0 i}$, $B^i \equiv - \frac{1}{2} \epsilon^{i k l} F_{k l}$, and $\pi \equiv \dot{\phi}$, is an averaged quantity over the box (we leave the sum over all sites and the division by the comoving volume implicit). We remark that quantities can be weighted to ensure that regions near strings are more relevant to the overall average. We will use $\mathcal{L}$ as the weight function.

From here we can use these definitions to define mean string separation and  velocity estimators from \cite{Correia:2019bdl, Correia:2020yqg}. Mean string separation measures are defined as,
\begin{align}\label{diagxi}
  \xi_{\mathcal{L}} = \sqrt{ \frac{\mathcal{V}}{l_{\mathcal{L}} } }\,,
  && 
  \xi_W = \sqrt{\frac{\mathcal{V}}{l_W}} \,,
\end{align}
where $\mathcal{V}$ is a box volume and $l_{\mathcal{L}} =-\mu(\tau)/\mathcal{L}$ is a length based on the Lagrangian density being strongly negatively peaked at string cores, with the tension being $\mu(\tau) = 2\pi \eta_0^2 a^{-2m}$. We also define $l_W = \frac{2}{3}n_p \delta x$ as a length measure, where $n_p$ is the number of cells of size $\delta x$ in the simulation box pierced by winding-one strings. Two different estimators to compute the root-mean-square velocity are defined as
\begin{align}\label{diagvel}
  v^2_{\phi} = \frac{2R}{1+R}\,,
  &&
  v^2_{\omega} = \frac{1}{2} \bigg( 1+\frac{3p}{\rho} \bigg)\,,
\end{align}
where $R = \rho_\pi / \rho_D$. These two estimators are expected to differ for the lattice spacing chosen above ($\delta x$) \cite{Hindmarsh:2017qff,Correia:2019bdl,Correia:2021tok}, with $v^2_\phi$ being lower than $v^2_\omega$ by roughly $10\%$ maximally. Reducing lattice spacing improves agreement between these two estimators \cite{Hindmarsh:2017qff,Correia:2021tok}.

\section{Analytical model of a string network with varying tension}
\label{III}

\subsection{Formulation of VOS}

Analytical models describing cosmic string networks with time-dependent tension, based on the NG approximation, have been studied in Refs.~\cite{Yamaguchi:2005gp, Revello:2024gwa, Chun:2025ret}. Let us derive the description of cosmic strings with time-dependent tension and clarify the meaning of the relevant parameters.

A zero-width string, whose worldsheet is parametrized by the spacetime coordinates $X^{\mu}(\zeta^a)$ and whose tension $\mu(\tau)$ depends on conformal time, is described by the NG action.
\begin{equation}
\label{NGaction}
    S = -  \int \mu(\tau) \sqrt{- \gamma} d^2 \zeta,
\end{equation}
where $\gamma$ is the determinant of the induced string world-sheet metric, which is parametrized by $\zeta^a$ and defined as $\gamma_{ab} \equiv X^{\mu}_{,a} X^{\nu}_{,b} g_{\mu \nu}$, where $a,b = 0,1$ are indices of the world-sheet coordinates. We choose the string world-sheet parametrization $\zeta^a = (\tau,\sigma)$ in such a way that
\begin{equation}
\label{Tgauge}
  X^0 = \tau, \qquad  \dot{\textbf{X}} \cdot \textbf{X}^{\prime} = 0,
\end{equation}
where $\dot{X} \equiv \frac{d X}{d \tau}$ and $X^{\prime} \equiv \frac{d X}{d \sigma}$.

Variation of $X^\mu(\zeta^a)$ in the action \eqref{NGaction}, together with the temporal orthogonal gauge \eqref{Tgauge}, leads to the equations of motion \cite{Emond:2021vts}.
\begin{equation}
\begin{gathered}
\label{MicroEqOfMot}
        \frac{\dot{\epsilon}}{\epsilon} + \dot{\textbf{X}}^2 \left( \frac{\dot{\mu}}{\mu} + 2 \frac{\dot{a}}{a}  \right) =0 , \\
        \ddot{\textbf{X}}+ \dot{\textbf{X}} \left( \frac{\dot{\epsilon}}{\epsilon} + \frac{\dot{\mu}}{\mu} + 2 \frac{\dot{a}}{a}  \right) = \frac{1}{\epsilon} \left( \frac{\textbf{X}^{\prime} }{\epsilon} \right)^{\prime},
\end{gathered}
\end{equation}
where $\epsilon^2 \equiv \frac{\textbf{X}^{\prime \, 2}}{1 - \dot{\textbf{X}}^2}$.

Following the procedure of Ref.~\cite{Martins:1996jp} for the VOS model, we introduce the macroscopic total energy $E$ of long strings and the rms velocity $v$ of the string network, defined as follows:
\begin{equation}
\begin{gathered}
    \label{AvVar}
    E \equiv \mu(\tau) a(\tau) \int \epsilon d \sigma, \quad
    v^2 \equiv \frac{\int \dot{\textbf{X}}^2 d \sigma }{\int \epsilon d \sigma} \equiv \left\langle \dot{\textbf{X}}^2 \right\rangle,
\end{gathered}
\end{equation}
where the integrals above represent integration over the string length and summation over all cosmic strings in the string network. Differentiating \eqref{AvVar} and substituting the first equation of motion \eqref{MicroEqOfMot}, we obtain the rate of change of the energy of the string network:
\begin{equation}
    \label{EnergEq}
    \frac{\dot{E}}{E} =  \frac{\dot{\mu}}{\mu} \left(1 - v^2\right) + \frac{\dot{a}}{a} \left(1-2 v^2\right) - \tilde{c} \frac{v}{L},
\end{equation}
where we have also introduced the loop chopping efficiency $\tilde{c}$ \cite{Martins:1996jp, Bennett:1985qt}, which accounts for the loss of energy of long strings due to loop production.

Multiplying the second equation of motion \eqref{MicroEqOfMot} by $\dot{\textbf{X}}$ and $\epsilon$, and averaging the resulting equation of motion over the string network, we obtain the following expression for the rms velocity
\begin{equation}
\label{VelEq}
    \frac{\dot{v}}{v} + \left(1 - v^2\right) \left[ \frac{\dot{\mu}}{\mu} + 2 \frac{\dot{a}}{a} \right] =  \frac{  \left\langle \frac{ \textbf{X}^{\prime \prime } \cdot \dot{\textbf{X}} }{\epsilon} \right\rangle }{ v^2} ,
\end{equation}
where we have made the assumption that $\left\langle \dot{\textbf{X}}^2 \right\rangle \left\langle \dot{\textbf{X}}^2 \right\rangle = \left\langle \dot{\textbf{X}}^4 \right\rangle$ (see Appendix of Ref.~\cite{Martins:2020jbq} for a discussion).

The form of the expression $\left\langle \frac{ \textbf{X}^{\prime \prime } \cdot \dot{\textbf{X}} }{\epsilon} \right\rangle$ can be determined by considering this product for some particular string NG configurations, as was done in Ref.~\cite{Martins:2000cs}, leading to
\begin{equation}
\begin{gathered}
    \label{MomentPar}
    \left\langle \frac{ \textbf{X}^{\prime \prime } \cdot \dot{\textbf{X}} }{\epsilon} \right\rangle \equiv \frac{v}{R} k(v) \underset{\mathrm{NG}}{=} \\ \underset{\mathrm{NG}}{=}  \frac{2 \sqrt{2}}{\pi} \frac{v  (1-v^2) (1+2\sqrt{2} v^3)}{R} \frac{1-8 v^6}{1+8v^6} ,
\end{gathered}
\end{equation}
where $R$ is an average curvature radius of the conformal string. 

To close the system of equations \eqref{EnergEq} and \eqref{VelEq}, one needs to make an additional assumption that the string network is Brownian on large scales \cite{Martins:1996jp}:
\begin{equation}
    E = V \frac{\mu}{a^2 \xi^2} \equiv \rho V,
\end{equation}
where $\xi$ is the conformal correlation length. 

Assuming that $R \sim \xi \sim L$ and taking into account the numerical factors in the definitions of $\tilde{c}$ and $k(v)$, one obtains the final system of equations:
\begin{equation}
\begin{gathered}
        \label{MacroEqOfMot}
    2 \frac{\dot{\xi}}{\xi} = v^2 \left( 2 \frac{\dot{a}}{a} + \frac{\dot{\mu}}{\mu} \right) + \tilde{c} \frac{v}{\xi}, \\
    \dot{v} = (1-v^2) \left[ \frac{ k(v)}{\xi} - v \left( \frac{\dot{\mu}}{\mu} + 2 \frac{\dot{a}}{a} \right) \right].
\end{gathered}
\end{equation}

The model \eqref{MacroEqOfMot} provides a NG description of the string network. A detailed comparison between field-theory simulations and the model \eqref{MacroEqOfMot} showed that, in order to reproduce the simulation results over a wide range of cosmological expansion rates, the model must be extended as follows \cite{Correia:2019bdl, Correia:2021tok}:
\begin{equation}
\begin{gathered}
        \label{MacroEqOfMotAH}
    2 \dot{\xi} = v^2 \xi \left( 2\frac{\dot{a}}{a} + \frac{\dot{\mu}}{\mu} \right) + F(v), \\
    \dot{v} = (1-v^2) \left[ \frac{k_{\rm AH}(v)}{\xi} - v \left( \frac{\dot{\mu}}{\mu} + 2 \frac{\dot{a}}{a} \right) \right],
\end{gathered}
\end{equation}
where the energy-loss function is given by
\begin{equation}
        \label{Floss}
     F(v) = \tilde{c} v + d \left( k_0 - k_{\rm AH}(v) \right)^r, 
\end{equation}
and the momentum parameter is
\begin{equation}
        \label{Kmoment}
     k_{\rm AH}(v) = k_0 \frac{1-(q v^2)^{\beta}}{1+ (q v^2)^{\beta}}.
\end{equation}
The parameters in Eqs.~\eqref{Floss}-\eqref{Kmoment} are phenomenological and are calibrated using numerical simulations. Their physical interpretation is discussed in Ref.~\cite{Martins:2016ois}. In particular, numerical simulations of field-theory cosmic string networks yielded the following parameter values \cite{Correia:2021tok}:\footnote{We use the parameter values corresponding to the lattice spacing $\delta x = 0.5$, as given in Table~I of Ref.~\cite{Correia:2021tok}.}
\begin{equation}
\begin{gathered}
    \label{ParameterFixed}
    \tilde{c}=0.23, \; d =0.36, \; r=1.72, \\
    \beta=2.5, \; k_0=1.06, \; q = 2.83 \\
    \text{and} \\
    \tilde{c}=0.0, \; d =0.39, \; r=1.36, \\
    \beta=2.32, \; k_0=1.18, \; q = 2.59
\end{gathered}
\end{equation}
for $v^2_{\phi}$ and $v^2_{\omega}$ estimators, respectively. 
Two sets of the above parameters, given in Eq.~\eqref{ParameterFixed}, arise from different methods of measuring the string velocity in simulations, namely, $v_{\omega}$ and $v_{\phi}$, as defined in Eq.~\eqref{diagvel}. We use both sets of parameters to investigate whether the scenario with varying tension can be described by this calibrated extension of the VOS model. The difference between the $v_{\phi}$ and $v_{\omega}$ measurements disappears when the lattice spacing is reduced to $\delta x = 0.25$ \cite{Correia:2021tok}. However, since our simulations are performed with $\delta x = 0.5$, we compare the varying-tension simulations with the calibrated VOS of the standard AH model performed at the same lattice spacing. This ensures a consistent, like-for-like comparison, in which possible differences are less likely to be affected by the numerical resolution. 

\subsection{Scaling solutions of VOS}

To determine the asymptotic scaling solutions of the string network described by \eqref{MacroEqOfMotAH} for an expansion law $a \propto \tau^n$, we assume a linear scaling regime in which $\xi=\varepsilon\tau$, with both $v$ and $\varepsilon$ being constants, denoted by $v_{\rm sc}$ and $\varepsilon_{\rm sc}$, respectively. Substituting this ansatz into \eqref{MacroEqOfMotAH}, we obtain the following algebraic system  that determines the asymptotic scaling values (a more general analysis of possible scaling solutions is presented in Ref.~\cite{Coelho:2026oeg}):
\begin{equation}
\begin{gathered}
\label{AlgebrSol}
 n v_{\rm sc}^2  (1 - m ) + \frac{F(v_{\rm sc})}{2 \varepsilon_{\rm sc}} = 1 , \quad
\frac{1}{2} = \frac{ n v \varepsilon_{\rm sc} ( 1 - m )}{k_{\rm AH}(v_{\rm sc})}.
\end{gathered}
\end{equation}

The stability of these scaling solutions can be investigated using linear perturbation theory. In particular, the system \eqref{MacroEqOfMotAH} can be rewritten in the following form:
\begin{equation}
\label{PertEq}
\frac{d  \textbf{Z}}{dy} = J(v_{\rm sc}, \varepsilon_{\rm sc})  \textbf{Z},
\end{equation}
where $ \textbf{Z} \equiv \left( \begin{matrix} \delta v \\ \delta \varepsilon \end{matrix} \right)$, $y \equiv \ln \tau$, and $J$ denotes the Jacobian matrix evaluated at the scaling solution $(v_{\rm sc},\varepsilon_{\rm sc})$.

Equation \eqref{PertEq} indicates that perturbations around a stable scaling solution decay as power laws, namely, $\delta \epsilon, \, \delta v \propto \tau^{\kappa_{1,2}}$, where $\kappa_{1,2}$ are the two eigenvalues of the Jacobian matrix. Stability requires both eigenvalues to be negative. During both the radiation- and matter-dominated epochs, the scaling solution remains stable until the rms velocity approaches the speed of light, $v_{\rm sc} \rightarrow 1$. This occurs at
\begin{equation}
m^* \equiv 1 - \frac{k_{\rm AH}(1)}{n [F(1) + k_{\rm AH}(1)]},
\end{equation}
which determines the maximum rate of tension change for which the string network can maintain a scaling regime. Detailed analyses of the VOS model calibrated to NG strings are discussed in Ref.~\cite{Revello:2024gwa}.
\section{Results of simulations and comparison with VOS}
\label{IV}
\begin{figure}[t!]
    \centering
    \includegraphics[width=1\linewidth]{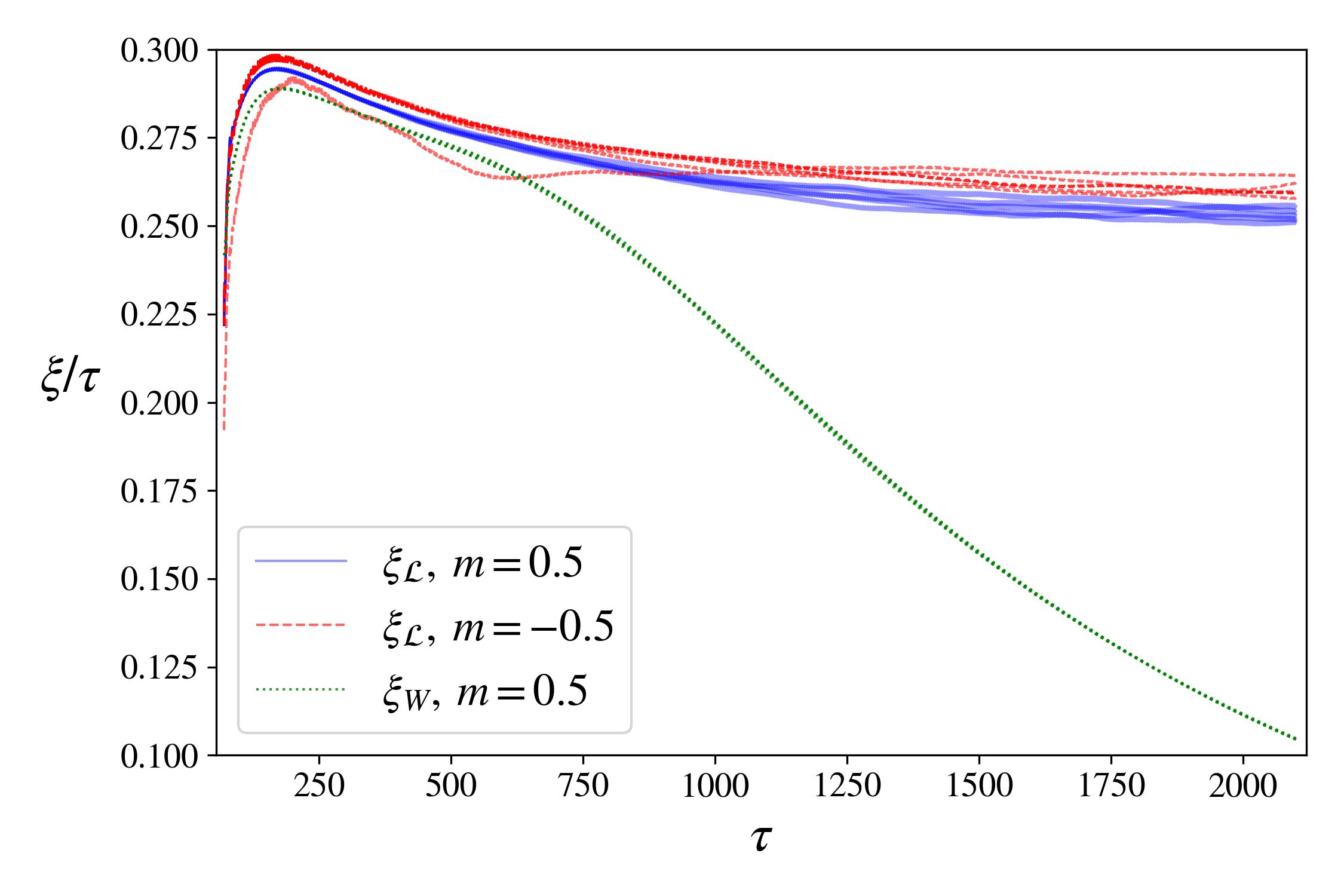}
    \includegraphics[width=1\linewidth]{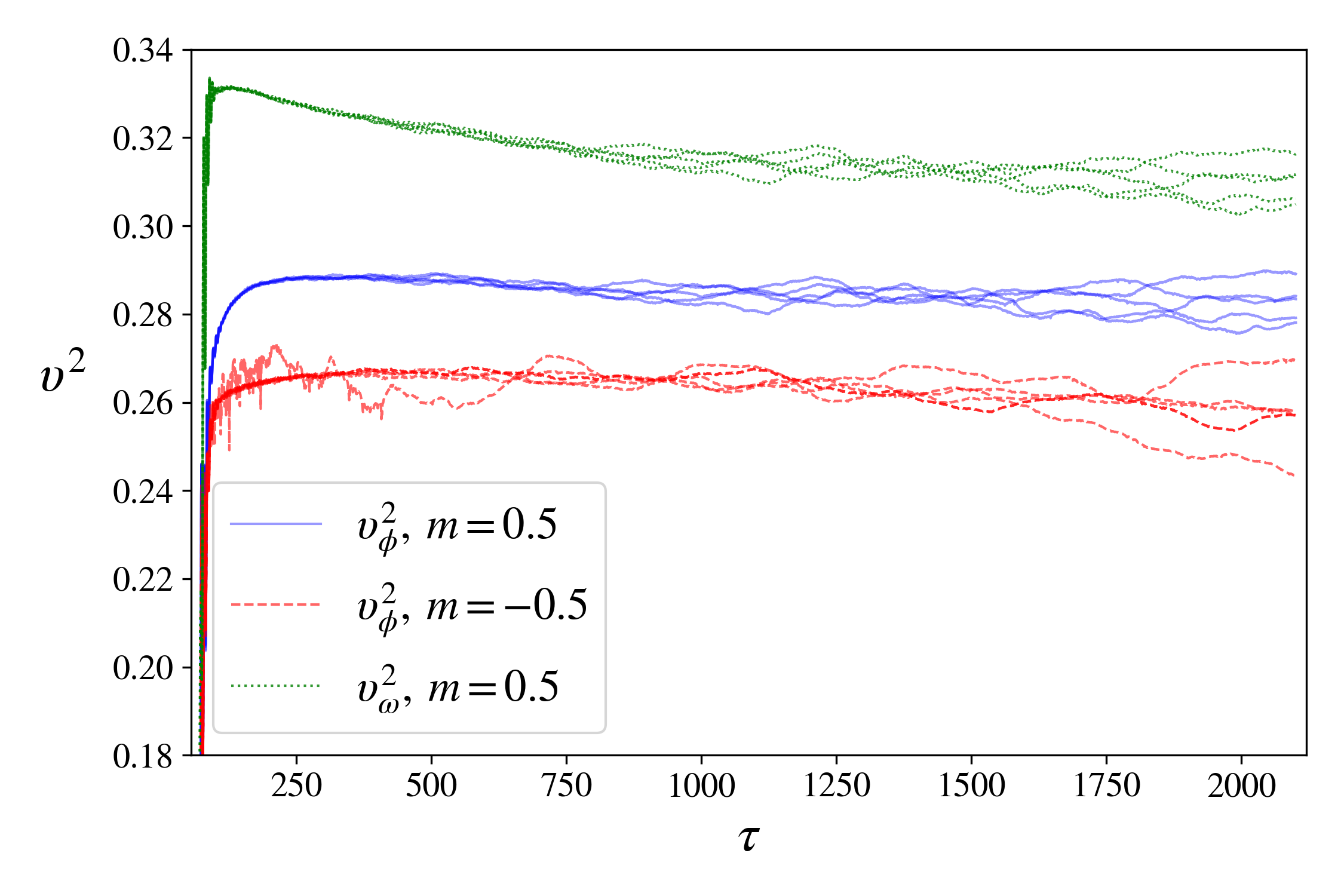}
    \caption{ \label{fig:ExampleSimScal} Examples of scaling behavior observed in numerical simulations during the radiation-dominated epoch for $m=0.5$ and $m=-0.5$. The top panel shows measurements of $\xi_{\mathcal{L}}$. For $m = 0.5$, $\xi_W$ shows no scaling regime. The potential barrier decreases faster than the vacuum field oscillations redshift, so the field does not settle into a vacuum and the strings cannot be identified reliably. The bottom panel shows examples of the scaling behavior of the rms velocities.}
\end{figure}

\begin{figure}[t!]
\centering
\includegraphics[width=0.48\textwidth]{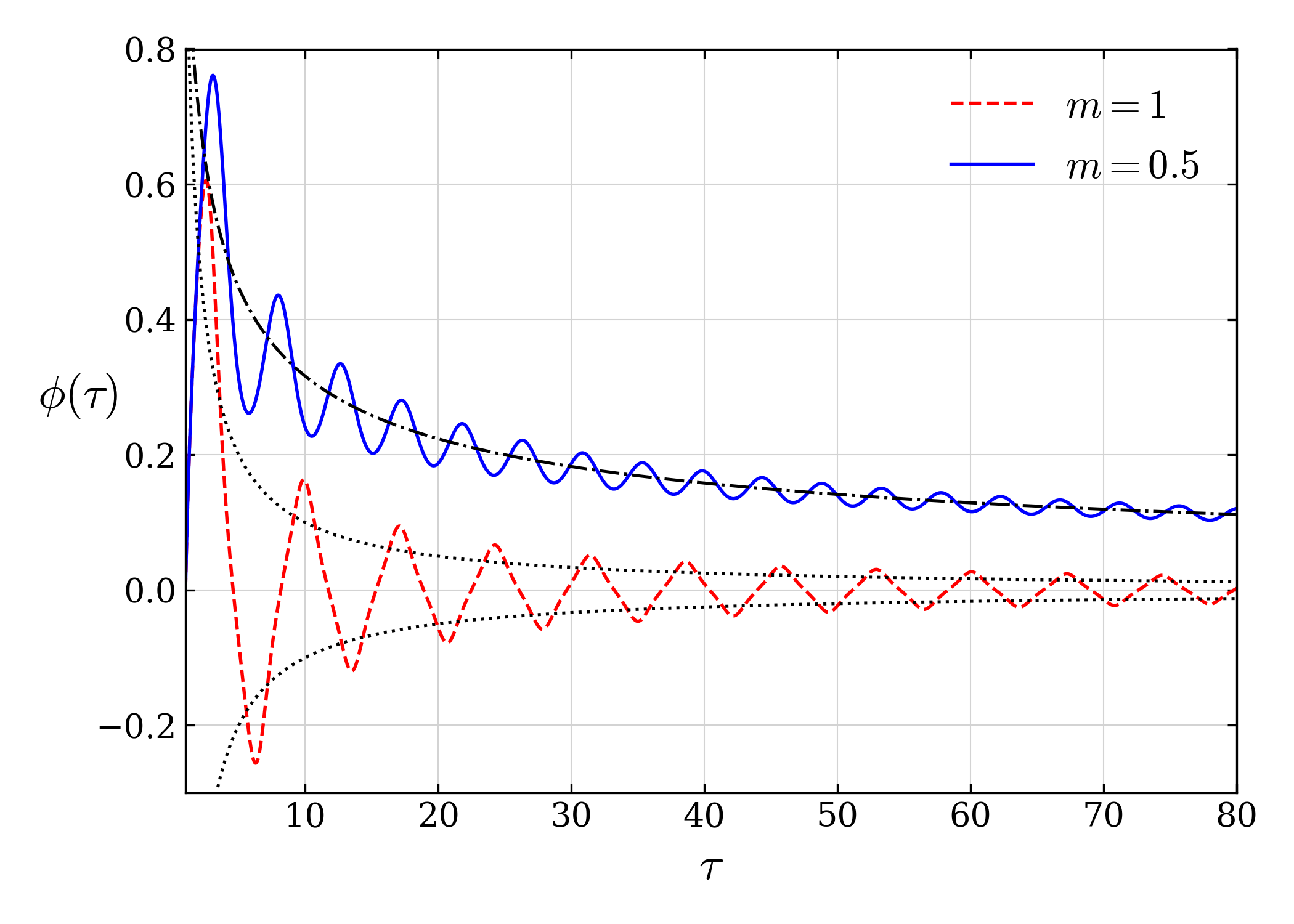}
\caption{Evolution of the radial field component for different values of $m$. The solid line shows the field evolution for $m=s=0.5$, while the dash-dotted line indicates the position of the potential minimum about which the field oscillates. The dashed line shows the evolution of the scalar field for $m=s=1$, starting from the same initial conditions. In this case, the potential decreases sufficiently rapidly that the field can acquire enough energy to cross the potential barriers and oscillate between the minima indicated by the dotted lines. \label{fig:Jumps} }
\end{figure}

We use the parameters obtained from the high-resolution calibration of the extended VOS model in Ref.~\cite{Correia:2021tok}, as given in Eq.~\eqref{ParameterFixed}. The perturbations around the scaling solutions are expected to decay as power laws, with the corresponding exponents governing the slower decay given in Table~\ref{tab:m_values}.

\begin{table}[ht]
    \centering
    \begin{tabular}{c|c|c|c|c|}
        \hline
        m & -0.5 & -0.25 &  0.25 & 0.5 \\
        \hline
        -$ \kappa_{\rm rad}$ & 1.10|1.06 & 1.08|1.04 & 1.06|1.02 & 1.04|1.01 \\
        -$\kappa_{\rm mat}$ & 1.16|1.18 & 1.14|1.13 &  1.10|1.06 & 1.07|1.03 \\
        \hline
    \end{tabular}
    \caption{ \label{tab:m_values} Values of the power-law exponents, given by the eigenvalues $\kappa$, that determine the slowest decay rate of perturbations for different values of $m$. The left and right columns correspond to the upper and lower sets of parameters in Eq.~\eqref{ParameterFixed}, respectively. }
\end{table}

In the next section, we use the power-law exponents from Table~\ref{tab:m_values} to obtain more accurate estimates of the asymptotic values from field-theory simulations, which we then compare with the predictions of the analytical VOS model given by Eqs.~\eqref{MacroEqOfMotAH}-\eqref{Kmoment}.

We performed numerical simulations of string network evolution during both the radiation-dominated epoch, when $a \propto \tau$, and the matter-dominated epoch, when $a \propto \tau^2$. We considered four values of the parameter $m$, which controls the rate of string tension variation: $m = \left[-0.5,\,-0.25,\,0.25,\,0.5\right]$. For each value of $m$, we performed several realizations with different initial conditions.

As described in Section~\ref{IIB}, we directly measure the rms velocities of strings, $v^2_{\phi}$ and $v^2_{\omega}$. Although the absolute values differ depending on the velocity measurement method, both techniques show that the string network approaches a scaling solution with a constant rms velocity. This behavior is illustrated in the lower panel of Fig.~\ref{fig:ExampleSimScal} for the radiation-dominated epoch.

We also measure the characteristic length using two different methods, namely $\xi_\mathcal{L}$ and $\xi_W$, as described in Section~\ref{IIB}. Similarly to the velocity measurements, the two methods yield somewhat different absolute values. Nevertheless, both measurements demonstrate that the string network approaches a scaling regime in which the characteristic length grows linearly with conformal time, $\xi \propto \tau$. Examples of scaling are shown in the top panel of Fig.~\ref{fig:ExampleSimScal} for the radiation-dominated epoch.

In field-theory simulations, the allowed values of $m$ are constrained by an additional physical mechanism. Since the potential barrier in the action \eqref{LagrangIn} decreases with time according to \eqref{DynamParam}, Hubble damping may become insufficient to bring the field to one of the potential minima. When the potential decreases faster than the field oscillations are redshifted due to the expansion of the Universe, the field can escape from a potential minimum and continue to cross the potential barriers throughout its subsequent evolution. A one-dimensional solution of the radial field equation, Fig.~\ref{fig:Jumps}, shows the mechanism directly. When the barrier decreases faster than the oscillation amplitude redshifts, the field crosses the barrier repeatedly and never settles into a vacuum. As a result, the distinction between strings and field oscillations becomes ambiguous, and the measurements obtained from the simulations no longer reliably describe the evolution of a string network. This limitation is reflected in the $\xi_W$ measurements during the radiation-dominated epoch for $m=0.5$, shown in Fig.~\ref{fig:ExampleSimScal}. Among all the simulations performed in this study, this is the only parameter choice for which we do not obtain reliable measurements. We take this limitation into account when interpreting the simulation results.

\begin{figure*}[p]
\centering
\includegraphics[width=0.45\textwidth]{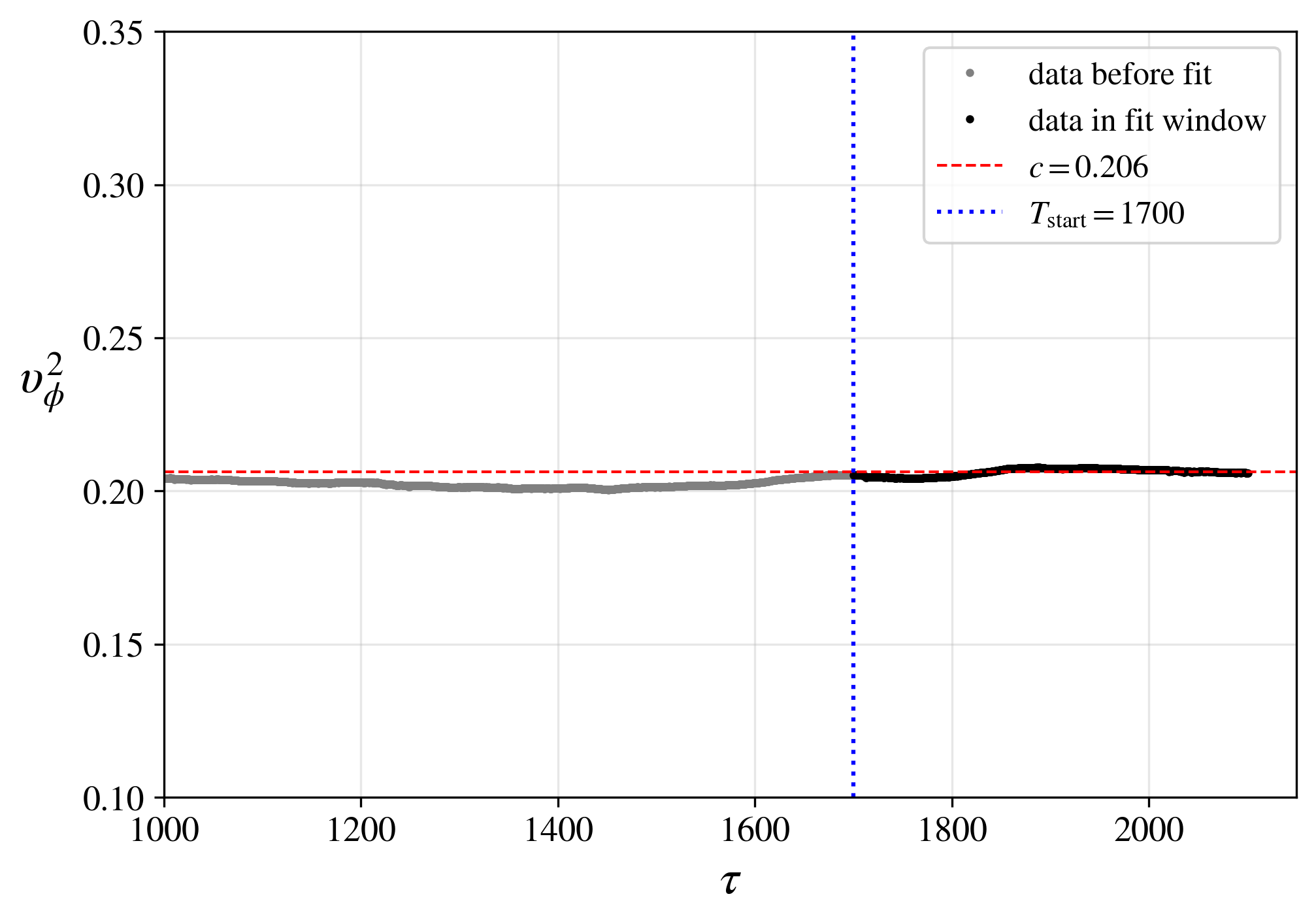}
\includegraphics[width=0.45\textwidth]{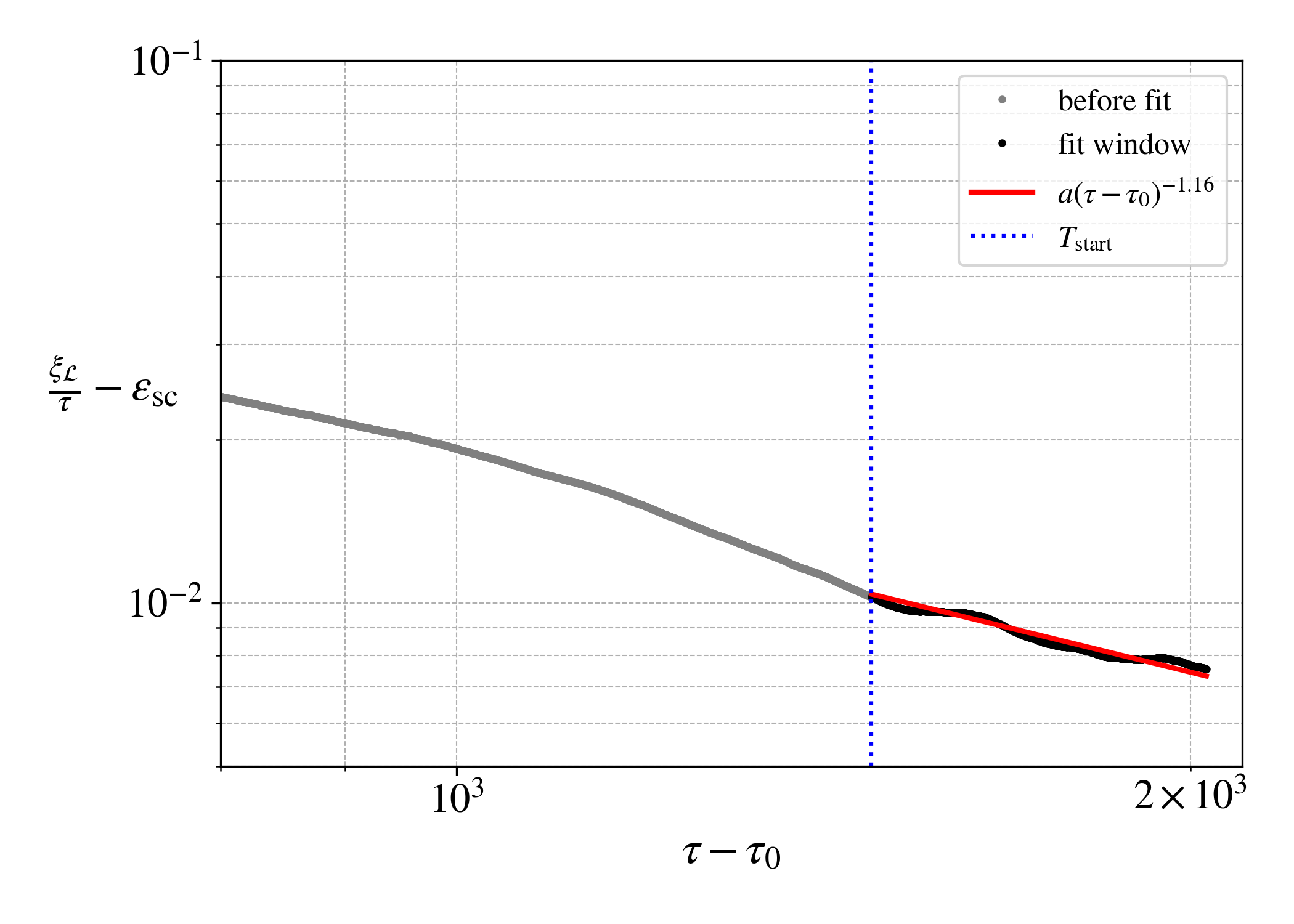}
\caption{Example of fitting the velocity $v^2_{\phi}$ and the characteristic length $\xi_{\mathcal{L}}/\tau$ as functions of time. In the first case, we use a simple constant function, while in the latter, we use a function of the form $\varepsilon_{\rm sc} + a \left( \tau-\tau_0 \right)^{-p}$. The constant $\varepsilon_{\rm sc}$ represents the asymptotic value approached by the decreasing power-law contribution predicted by the VOS model. For this particular case, corresponding to a matter-dominated period ($n=2$) with a varying tension ($m=-0.5$), the power-law exponent is $p=1.16$, as predicted by the VOS model. We set the offset to $\tau_0=70$. However, varying the offset primarily affects the parameter $a$ and has only a small influence on the asymptotic value $\varepsilon_{\rm sc}$.\label{fig:fitting} }
\end{figure*}

\begin{figure*}[p]
\centering
\includegraphics[width=0.45\textwidth]{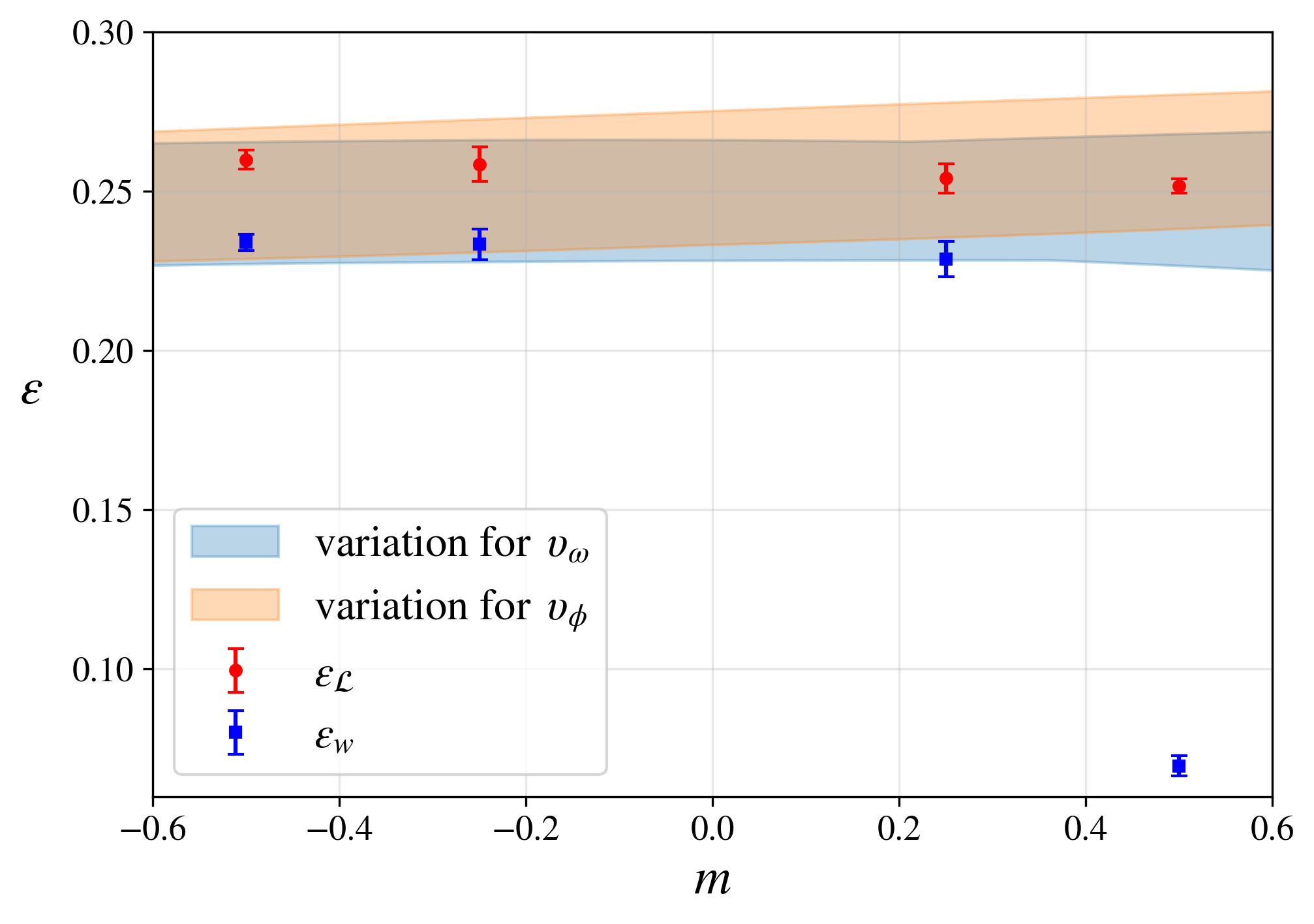}
\includegraphics[width=0.45\textwidth]{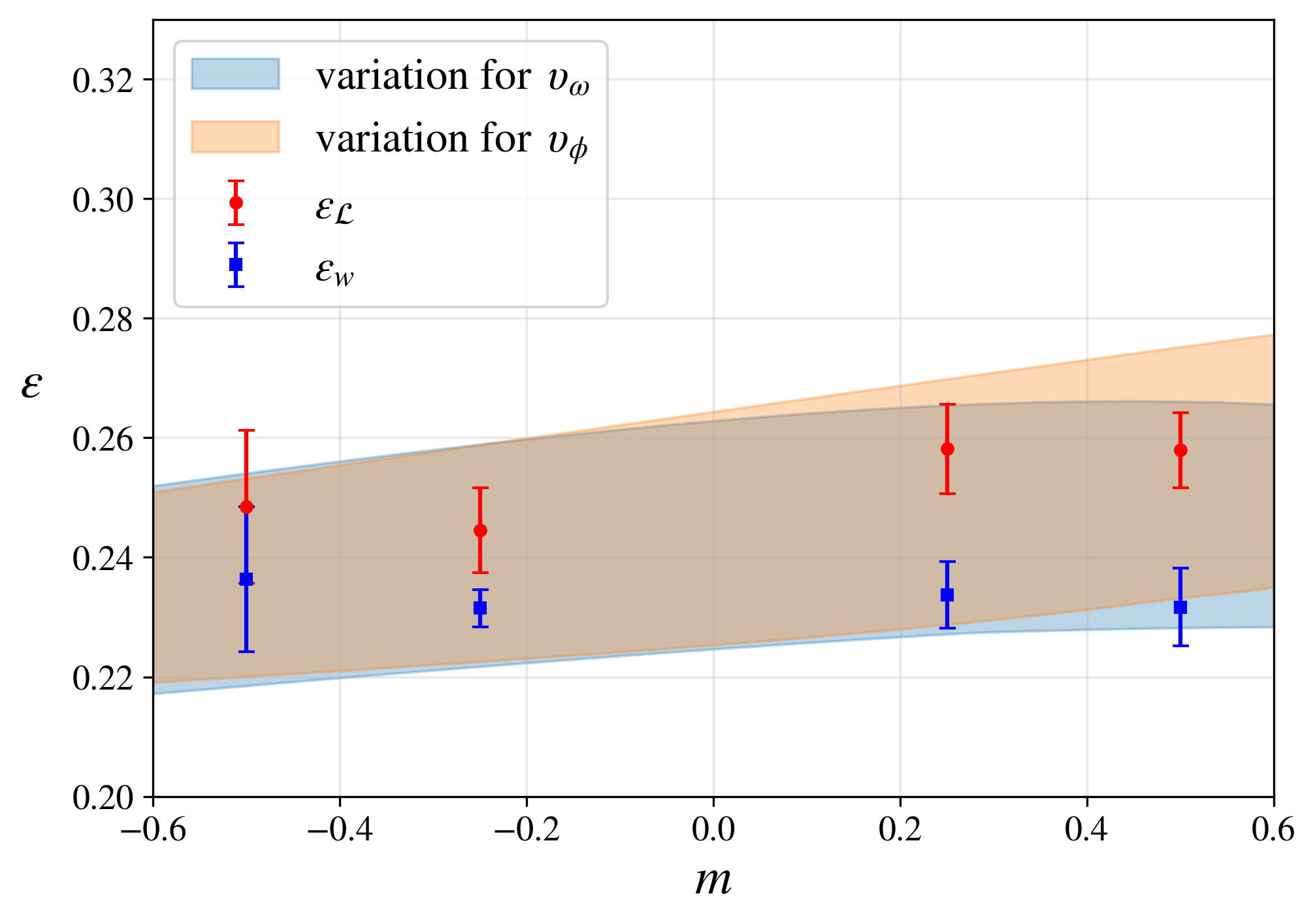}
\includegraphics[width=0.45\textwidth]{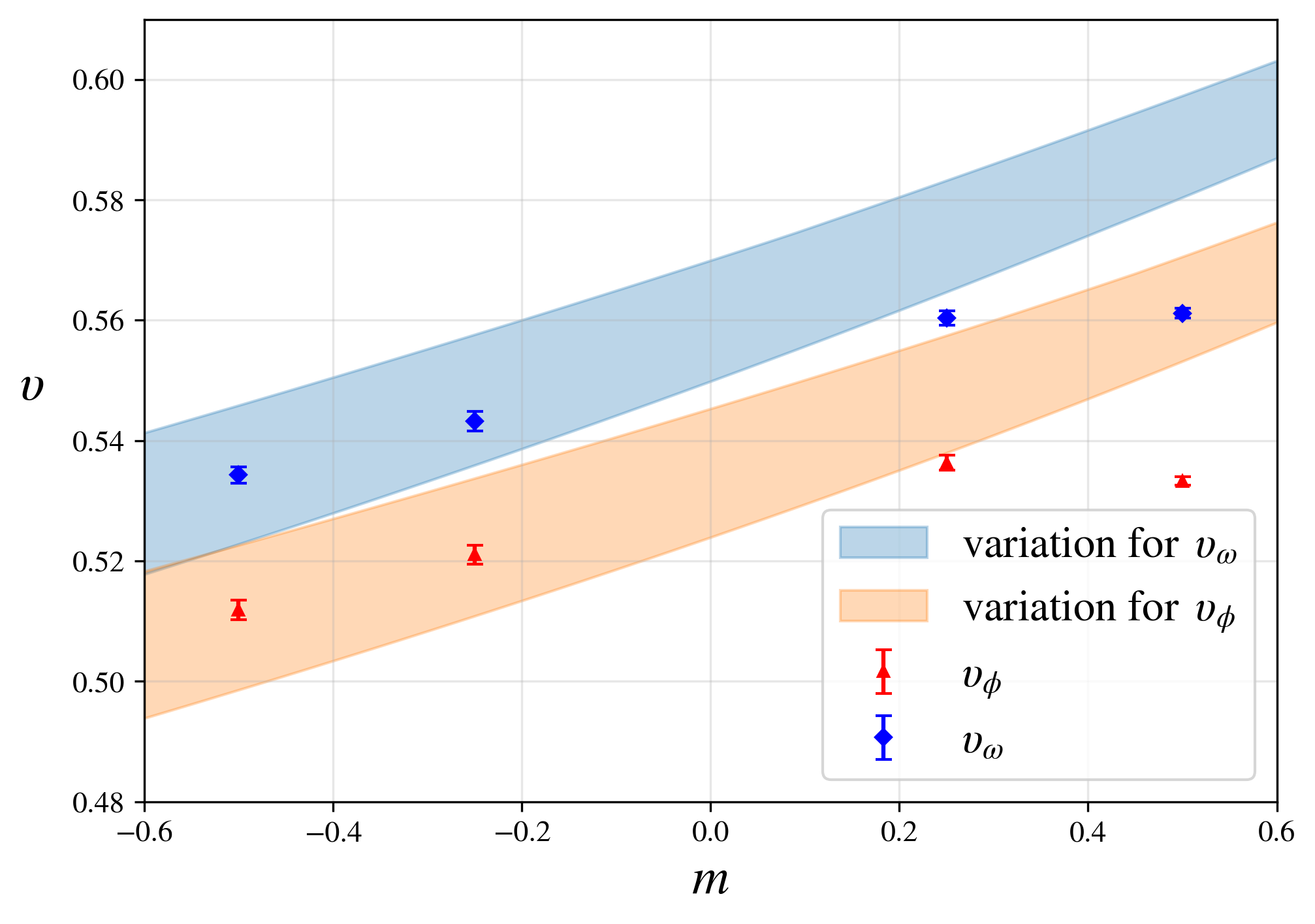}
\includegraphics[width=0.45\textwidth]{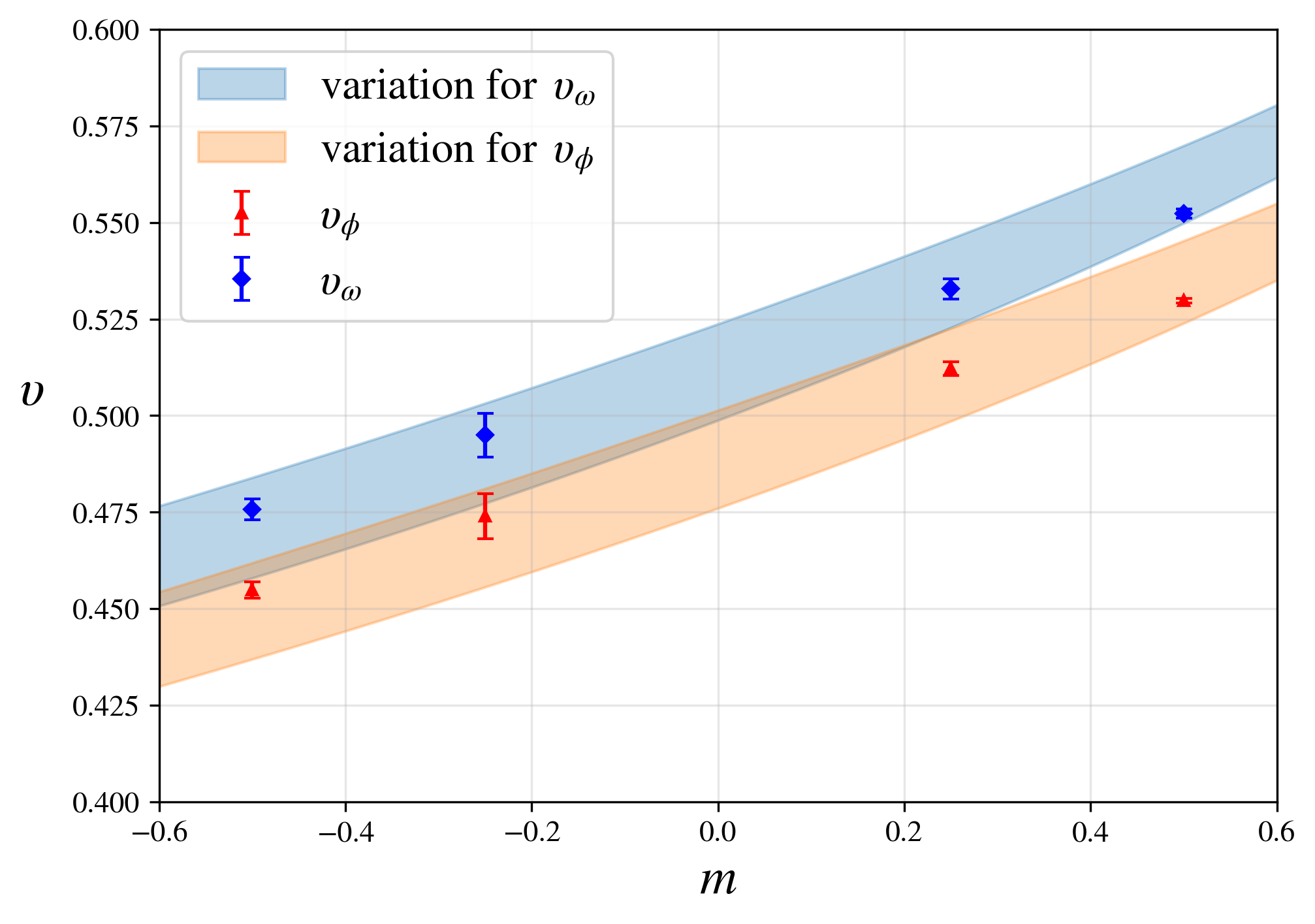}
\caption{  The shaded areas show the predictions of the extended VOS model with varying tension, whose parameters and associated uncertainties are calibrated in Ref.~\cite{Correia:2021tok} for $\delta x = 0.5$, as given in Table~I. These predictions are compared with field-theory simulations with dynamically varying string tensions. The corresponding quantities measured in the simulations are shown by dots: $\varepsilon_{\mathcal{L}}$, $\varepsilon_w$, $\upsilon_{\phi}$, and $\upsilon_{\omega}$. The left panels demonstrate results for the radiation-dominated epoch, and right panels for matter-dominated epoch.  \label{fig:Comparison} }
\end{figure*}

To quantify the asymptotic scaling behavior of the string network, we fit the numerical results with a decaying power-law function with a constant offset, as illustrated in Fig.~\ref{fig:fitting}. The power-law exponents used in the fits are taken from Table~\ref{tab:m_values}. We perform the fits for each simulation and account for the statistical uncertainties across different realizations. The resulting best-fit values and their corresponding uncertainties are shown in Fig.~\ref{fig:Comparison}.

In Fig.~\ref{fig:Comparison}, we see that the characteristic length $\xi$ changes only slightly with $m$ during both radiation- and matter-dominated epochs. The results of the field-theory simulations remain in good agreement with the calibrated extended VOS model. The variation in the rms velocity across different values of $m$ is more pronounced, and this behavior is also consistent with the predictions of the extended VOS model.

We should note that, during the radiation-dominated epoch, the simulation results for $m=0.5$ are less reliable due to large field oscillations in the vacuum. In contrast, during the matter-dominated epoch, all simulations maintain well-defined string network configurations. We therefore find good agreement between the field-theory simulations of strings with time-varying tension and the predictions of the extended VOS model in the matter-dominated epoch.

\section{Conclusion}

In this study, we carried out the first field-theory simulations of cosmic strings with varying tension. For this purpose, we modified the AH model given by Eq.~\eqref{LagrangIn} so that the potential becomes time-dependent in an expanding Universe. In particular, we chose the parameters of the potential to have the time dependence described by Eq.~\eqref{DynamParam}. This automatically leads to a growing string core, as described by Eq.~\eqref{StingCore}, which allows us to keep the cosmic string core resolvable throughout the entire timescale of the numerical simulations. Thus, this simple modification of the AH model allows us to simulate cosmic strings with time-varying tension, obeying $\mu \propto a^{-2m}$, without compromising the resolution of the simulations.

We carried out numerical simulations for $m=-0.5,-0.25,0.25,0.5$ in both radiation- and matter-dominated Universes. The numerical simulations for these varying-tension scenarios demonstrate the standard scaling behavior, as can be seen in Fig.~\ref{fig:ExampleSimScal}. However, when the amplitude of the field oscillations redshifts more slowly than the potential barrier decreases in the AH model, vacuum oscillations prevent us from properly identifying the string locations, which eventually spoils the measurement of the network parameters, as seen in Fig.~\ref{fig:ExampleSimScal} for $m=0.5$ in the radiation-dominated epoch. Unfortunately, this limitation of the model does not allow us to probe the scenario in which cosmic string loops can start to grow in an expanding Universe, as suggested by Ref.~\cite{Conlon:2024uob}. Nevertheless, for all other parameters choices, we obtain the standard scaling behavior of the network, which provides valuable information to testing the predictions of the VOS model.

The VOS model predicts scaling behavior for string networks with time-dependent tension that evolves as a power law of the scale factor, as discussed previously in the literature \cite{Yamaguchi:2005gp, Emond:2021vts, Revello:2024gwa}. Our numerical simulations of the field-theory model for strings with varying tension confirm this prediction, namely, that $\xi \propto \tau$ and $v \propto \text{const}$. However, the decay of the string energy density is modified due to the variation of the string mass per unit length: $\rho \propto \frac{a^{-2(m+1)}}{\tau^2}$.

The presence of a scaling regime is particularly noteworthy in the context of the field theory. When the string tension varies due to changes in the potential, it also affects the masses of the scalar and vector particles in the vacuum. The time dependence of the scalar and vector masses can, in principle, lead to changes in the efficiency of massive radiation from cosmic strings. Since we obtain scaling behavior for the string networks, this may tentatively suggest that the efficiency of massive radiation emission from long strings is not strongly affected by changes in the masses of the radiated particles. Clearly, further study is necessary to clarify the energy-loss mechanisms of long string networks in scenarios with varying string tension.

We extracted the asymptotic values from each simulation, considering different initial conditions, rates of string tension variation, and rates of Universe expansion. To determine these asymptotic values, we fitted the numerical results using a power-law decay of perturbations around the scaling solution predicted by the VOS model. This allowed us to probe $8$ different string network configurations, which we then compared with the predictions of the extended VOS model calibrated using simulations of the standard AH model in Ref.~\cite{Correia:2021tok}. With the exception of the radiation-dominated case with $s=m=0.5$, in which cosmic strings are not properly defined, we find good evidence that the VOS model provides accurate predictions for the network evolution in all the other configurations considered. Nevertheless, it would be interesting to extend the range of parameters governing the tension variation and Universe expansion in future studies, in order to explore regimes in which the deviations from the standard AH model become more significant.

We do not anticipate significant variation in the asymptotic scaling values of the string network for scenarios with $m \neq s$. When we change only the parameter $s$ parameter, while keeping $m$ fixed, it corresponds to a transformation from fat strings to physical strings \cite{Hindmarsh:2017qff,Correia:2020yqg}.However, further investigation of this question is required.

The results obtained in our study are relevant for making accurate predictions of the observational signatures of cosmic strings with varying tension. So far, all observational predictions, see Refs.~\cite{Emond:2021vts, Revello:2024gwa, Conlon:2025mqt, Brunelli:2026ndh, Avgoustidis:2026zck}, have relied solely on the VOS model calibrated using NG simulations. In the case of field-theory realizations, time-varying cosmic string tension may lead to additional observational signatures associated with massive particle radiation. In particular, one can anticipate modifications to the gravitational-wave spectrum cut-off associated with the Higgs mass, as well as changes to the Big Bang nucleosynthesis constraints, which have so far been studied only in the context of the standard AH model in Refs.~\cite{SantanaMota:2014xpw, Auclair:2019jip}. These possibilities provide interesting avenues for further exploration.

\section*{Acknowledgments}
JC acknowledges fruitful discussions with Carlos Martins and Mark Hindmarsh on this topic. JC (ORCID ID 0000-0002-3375-0997) acknowledges support from the European Union’s Horizon Europe research and innovation programme under the Marie Skłodowska-Curie grant agreement No. 101126636; IR (ORCID ID 0000-0001-8501-8629) is supported by IBS under the project code IBS-R018-D3.
We acknowledge EuroHPC for awarding project access (EHPC-REG-2024R02-178) to the EuroHPC supercomputer Leonardo, hosted by CINECA (ITALY), where these simulations were performed. 

\bibliography{Biblio}

\end{document}